\documentclass{aa}  

\usepackage{graphicx}
\usepackage{txfonts}
\begin{document}

   \title{Mean plane of the Kuiper belt beyond 50 AU in the presence of Planet 9}
   


   \author{Jian Li
          \inst{1}
          \and
          Zhihong Jeff Xia\inst{2} 
          }

   \institute{School of Astronomy and Space Science \& Key Laboratory of Modern Astronomy and Astrophysics in Ministry of Education, Nanjing University, 163 Xianlin Road, Nanjing 210023, PR China\\
              \email{ljian@nju.edu.cn}
         \and
             Department of Mathematics, Northwestern University, 2033 Sheridan Road, Evanston, IL  60208, USA\\
             \email{xia@math.northwestern.edu}
             }

   \date{Received September 15, 1996; accepted March 16, 1997}

 
  \abstract
   {A recent observational census of Kuiper belt objects (KBOs) has unveiled anomalous orbital structures. This has led to the hypothesis that an additional $\sim5-10~m_{\oplus}$ planet exists. This planet, known as Planet 9, occupies an eccentric and inclined orbit at hundreds of astronomical units. However, the KBOs under consideration have the largest known semimajor axes at  $a>250$ AU; thus they are very difficult to detect.}
   {In the context of the proposed Planet 9, we aim to measure the mean plane of the Kuiper belt at $a>50$ AU. In a comparison of the expected and observed mean planes, some constraints would be put on the mass and orbit of this undiscovered planet.}
    {We adopted and developed the theoretical approach of \citet{volk17} to the relative angle $\delta$ between the expected mean plane of the Kuiper belt and the invariable plane determined by the eight known planets. Numerical simulations were constructed to validate our theoretical approach. Then similar to \citet{volk17}, we derived the angle $\delta$ for the real observed KBOs with $100<a<200$ AU, and the measurement uncertainties were also estimated. Finally, for comparison, maps of the theoretically expected $\delta$ were created for different combinations of possible Planet 9 parameters.}
   {The expected mean plane of the Kuiper belt nearly coincides with the said invariable plane interior to $a=90$ AU. But these two planes deviate noticeably from each other at $a>100$ AU owing to the presence of Planet 9 because the relative angle $\delta$ could be as large as $\sim10^{\circ}$. Using the $1\sigma$ upper limit of $\delta<5^{\circ}$ deduced from real KBO samples as a constraint, we present the most probable parameters of Planet 9: for mass $m_9=10~m_{\oplus}$, orbits with inclinations $i_9=30^{\circ}$, $20^{\circ}$, and $15^{\circ}$ should have semimajor axes $a_9>530$ AU, 450 AU, and 400 AU, respectively; for $m_9=5~m_{\oplus}$, the orbit is $i_9=30^{\circ}$ and $a_9>440$ AU, or $i_9<20^{\circ}$ and $a_9>400$ AU. In this work, the minimum $a_9$ increases with the eccentricity $e_9$ ($\in[0.2, 0.6]$) but not significantly.}
   {}


   \maketitle
%

\section{Introduction}

The Kuiper belt is a disk of icy minor planets extending outward from the orbit of Neptune (i.e., $\sim30$ AU). At the time of writing, more than 600 distant Kuiper belt objects (KBOs hereafter) have been discovered at semimajor axes $a$ beyond 50 AU  up to more than 1000 AU. These objects can provide new and rich information on the dynamical environment of the very edge of the solar system. Especially, the farthest components with $a>250$ AU, including Sedna and 2012 VP113, exhibit clustering in the argument of perihelion and the longitude of the ascending node \citep{truj14, baty16}. One possible explanation for this orbital distributions is the existence of an additional $5-10~m_{\oplus}$ planet residing on an extremely wide orbit ($a_9=400-800$ AU), which has eccentricity $e_9=0.2-0.6$ and inclination $i_9=15^{\circ}-30^{\circ}$ \citep{baty16, brow16, baty19}. This hypothesized planet is currently referred to as Planet 9.

Moreover, \citet{bail16}, \citet{lai16}, and \citet{gome17} found that the secular perturbation from Planet 9 could induce a slow precession of the invariable plane of the solar system. In this work, the invariable plane is defined as the plane perpendicular to the total angular momentum vector of the eight known planets, and hereafter is named IP8. Over a timespan of $\sim4.5$ Gyr, the proposed Planet 9 can yield the current $6^{\circ}$ tilt between the Sun's equator and IP8. Then, we speculate that this mechanism could also cause the change of the apparent orbital planes of the distant KBOs.

The first attempt to determine the mean plane of the Kuiper belt was made more than a decade ago by \citet{coll03} and \citet{brow04}. As a much larger number of KBOs have been discovered in recent years, \citet{volk17} extended the study of the mean plane as a function of the semimajor axis. These authors found that, for the classical Kuiper belt between 42 AU $<a<$ 48 AU, the expected mean plane  has an inclination $\lesssim2^{\circ}$ with respect to IP8 from secular theory; this value is consistent with the current observations. At a larger semimajor axis $a>50$ AU, the expected mean plane is very flat and close to IP8 in the unperturbed solar system model (i.e., without any additional planet). Even by including the possible Planet 9 on high-inclination orbit, for the Kuiper belt exterior to but not too far away from 50 AU, this plane would be nearly unaffected and still confined to the local Laplacian plane determined by the Jovian planets (almost equivalent to IP8). But if the KBOs are approaching the orbit of Planet 9 at hundreds of AU, the associated mean plane could substantially deviate from IP8. \citet{laer19} found that from 50 AU to 150 AU the mean plane of the observed Kuiper belt is consistent with IP8, while \citet{baty19} reported that the orbital planes of the 14 known KBOs with $a>250$ AU are clustered around a common plane (at the 96.5\% confidence), which is inclined to the ecliptic by $\sim7^{\circ}$. In this paper, we aim to quantify exactly how far and how large the warp of the mean plane of the Kuiper belt can be produced by the perturbations from the unseen Planet 9. If such a warp is detectable, it could put constraints on the proposed mass and orbital elements of Planet 9. 

The rest of this paper is organized as follows. In Section 2, we develop both the theoretical and numerical approaches to determine the expected mean plane of the Kuiper belt beyond 50 AU, in the presence of Planet 9. Accordingly, we calculate the relative angle $\delta$ between this plane and IP8. In Section 3, we measure the angle $\delta$ for the currently observed KBOs with semimajor axes in the range $100-200$ AU and the measurement uncertainty is also evaluated. In Section 4, based on the analysis of the expected and measured values of $\delta$, we present the most probable parameters of Planet 9. The conclusions are summarized in Section 5.

\section{Expected mean plane}
\label{EMP}

\subsection{Theoretical approach}
\label{theory}

To determine the expected mean plane of the Kuiper belt, we consider the gravitational perturbations from the eight known planets plus a hypothetical Planet 9. As for the motion of a test KBO, the inclination vector is defined by $(q, p)=(\sin i \cos \Omega, \sin i \sin \Omega)$, where $i$ and $\Omega$ are the inclination and the longitude of ascending node, respectively. Throughout this paper, the reference plane is given as IP8. Then, according to the classical Laplace-Lagrange secular theory, the forced inclination vector $(q_0, p_0)$ can be written as \citep[see][chap.~7]{murr99}
\begin{eqnarray}
  &&q_0=-\sum_{i=1}^{9}\frac{\mu_i}{B-f_i}\cos(f_i t + \gamma_i),\nonumber\\
  &&p_0=-\sum_{i=1}^{9}\frac{\mu_i}{B-f_i}\sin(f_i t + \gamma_i),
\label{forced}
\end{eqnarray}
where $f_i$ and $\gamma_i$ are the secular nodal eigenfrequency and associated phase, respectively; and
\begin{eqnarray}
  &&\mu_i=+n\frac{1}{4}\sum_{j=1}^{9}I_{ji}\frac{m_j}{m_{\odot}}{\alpha_j}{\bar{\alpha}_j}b_{3/2}^{(1)}({\alpha_j}),\nonumber\\
  &&B=-n\frac{1}{4}\sum_{j=1}^{9}\frac{m_j}{m_{\odot}}{\alpha_j}{\bar{\alpha}_j}b_{3/2}^{(1)}({\alpha_j}),
\label{coefficient}
\end{eqnarray}
where $m_{\odot}$ is the mass of the Sun, $m_j$ is the mass of the $j$-th planet, $n$ is the mean motion of the test KBO, $I_{ji}$ is the amplitude corresponding to $f_i$; and
\begin{equation}
\alpha_j=\left\{
             \begin{array}{lr}
              a_j/a~~~~~~\mbox{if}~~a_j < a, &  \\
              a/a_j~~~~~~\mbox{if}~~a_j > a, &  
             \end{array}
\right.
\end{equation}
\begin{equation}
\bar{\alpha}_j=\left\{
             \begin{array}{lr}
              1~~~~~~~~~~~\mbox{if}~~a_j < a, &  \\
              a/a_j~~~~~~\mbox{if}~~a_j > a, &  
             \end{array}
\right.
\end{equation}
where $a$ and $a_j$ are the semimajor axes of the test KBO and the $j$-th planet, respectively. In Eq. (\ref{coefficient}), the Laplace coefficient is given by
\begin{equation}
b_{3/2}^{(1)}({\alpha})=\frac{1}{\pi}\int_{0}^{2\pi}\frac{\cos\psi \mbox{d}\psi}{(1-2\alpha\cos\psi+{\alpha}^2)^{3/2}}.
\label{laplace}
\end{equation}

Since the secular modes of the known planets are supposed to be unaffected by the distant Planet 9, \citet{volk17} express the forced inclination vector (Eq. (\ref{forced})) in an approximate form  
\begin{eqnarray}
  &&(q_0, p_0)\approx (q_0^0, p_0^0)+(q_0^1, p_0^1),\nonumber\\
  &&(q_0^1, p_0^1)=\frac{\mu_9}{-B+f_9}\left(\cos(f_9 t + \gamma_9), \sin(f_9 t + \gamma_9)\right),
\label{approximate}
\end{eqnarray}
where the subscript 9 refers to Planet 9; and
\begin{equation}
\mu_9\approx +n\frac{1}{4}\frac{m_9}{m_{\odot}}{\alpha_9}{\bar{\alpha}_9}b_{3/2}^{(1)}({\alpha_9})\sin i_9,
\label{mu9}
\end{equation}
\begin{equation}
f_9 \approx -n_9\frac{1}{4}\sum_{j=1}^{8}\frac{m_j}{m_{\odot}}\left(\frac{a_j}{a_9}\right)b_{3/2}^{(1)}\left(\frac{a_j}{a_9}\right),
\label{f9}
\end{equation}
where $n_9$ is the mean motion of Planet 9. \citet{volk17} also point out that, for $a>50$ AU, the vector $(q_0^0, p_0^0)$ represents the forced plane determined by the known plants, which should be nearly coincident with IP8. Therefore, the tilt of the forced plane of test KBOs in this $a$ region is solely induced by Planet 9 and  can be measured by the forced inclination
\begin{equation}
i_0=\sqrt{\left(q_0^1\right)^2+\left(p_0^1\right)^2}=\arcsin\left(\frac{\mu_9}{-B+f_9}\right).
\label{i0}
\end{equation}
In this paper, we consider the Kuiper belt extended from 50 AU all the way out to the neighborhood of Planet 9 at hundreds of AU; that is, the semimajor axis ratio $\alpha_9=a/a_9$ could not always be close to 1. Then unlike in \citet{volk17}, for the calculation of the Laplace coefficient $b_{3/2}^{(1)}({\alpha})$, we evaluate the integral in Eq. (\ref{laplace}) without any approximation.

Bearing in mind that, the proposed Planet 9 has a substantial eccentricity $e_9$. The value of $e_9$ is very important for determining the forced inclination $i_0$, as we see below. Borrowing the method from \citet{gome06} to implement the averaged effect of the eccentric Planet 9, we assume a scaled semimajor axis
\begin{equation}
\tilde{a}_9=a_9\sqrt{1-e_9^2}.
\end{equation}
Substituting the expression of $\tilde{a}_9$ for $a_9$ into Eqs. (\ref{mu9}) and (\ref{f9}), we can finally obtain the intrigue value of $i_0$ from Eq.  (\ref{i0}).

\begin{figure}
  \centering
  \begin{minipage}[c]{0.5\textwidth}
  \centering
  \hspace{0cm}
  \includegraphics[width=8.5cm]{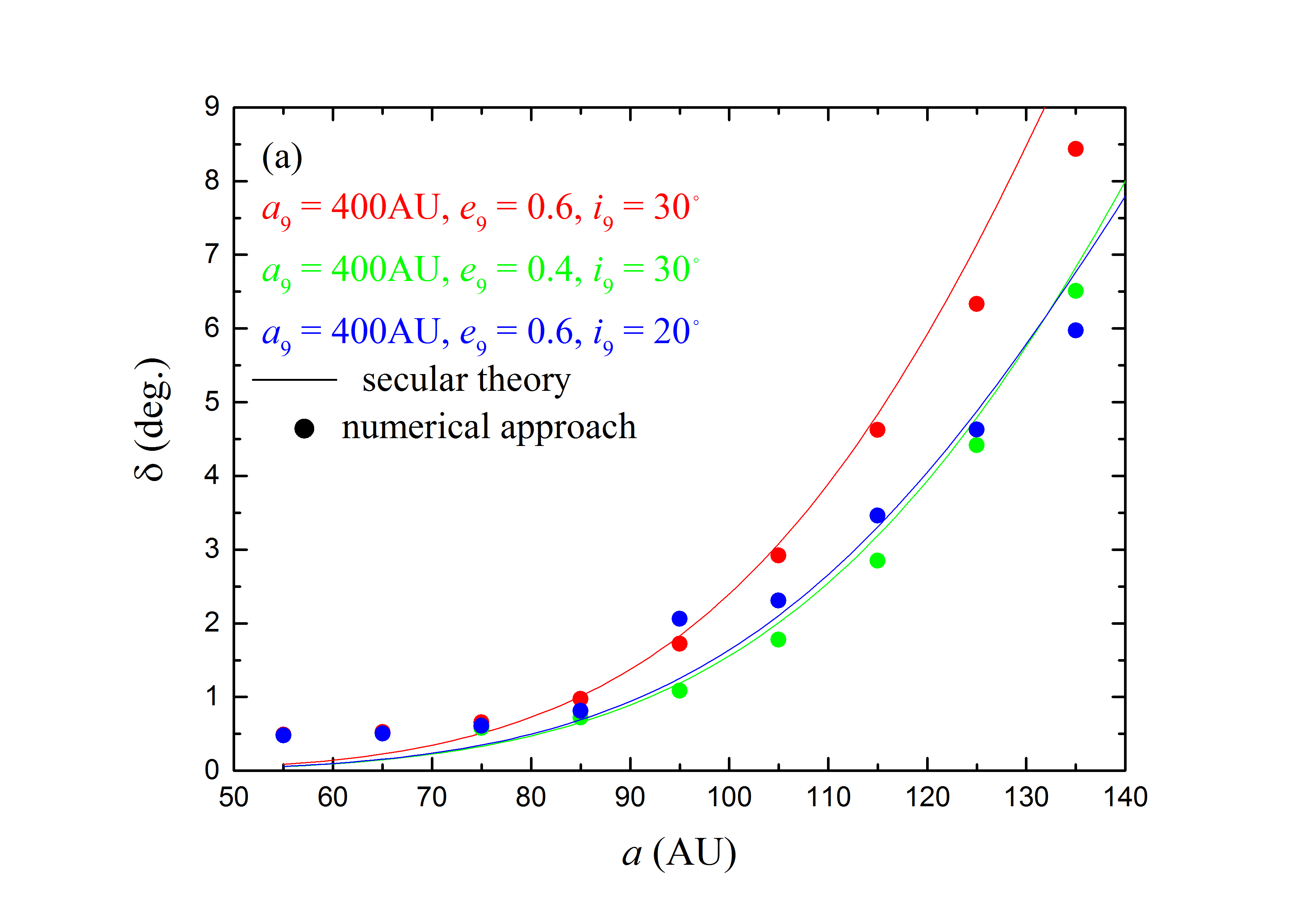}
  \end{minipage}
  \begin{minipage}[c]{0.5\textwidth}
  \centering
  \vspace{0cm}
  \includegraphics[width=8.5cm]{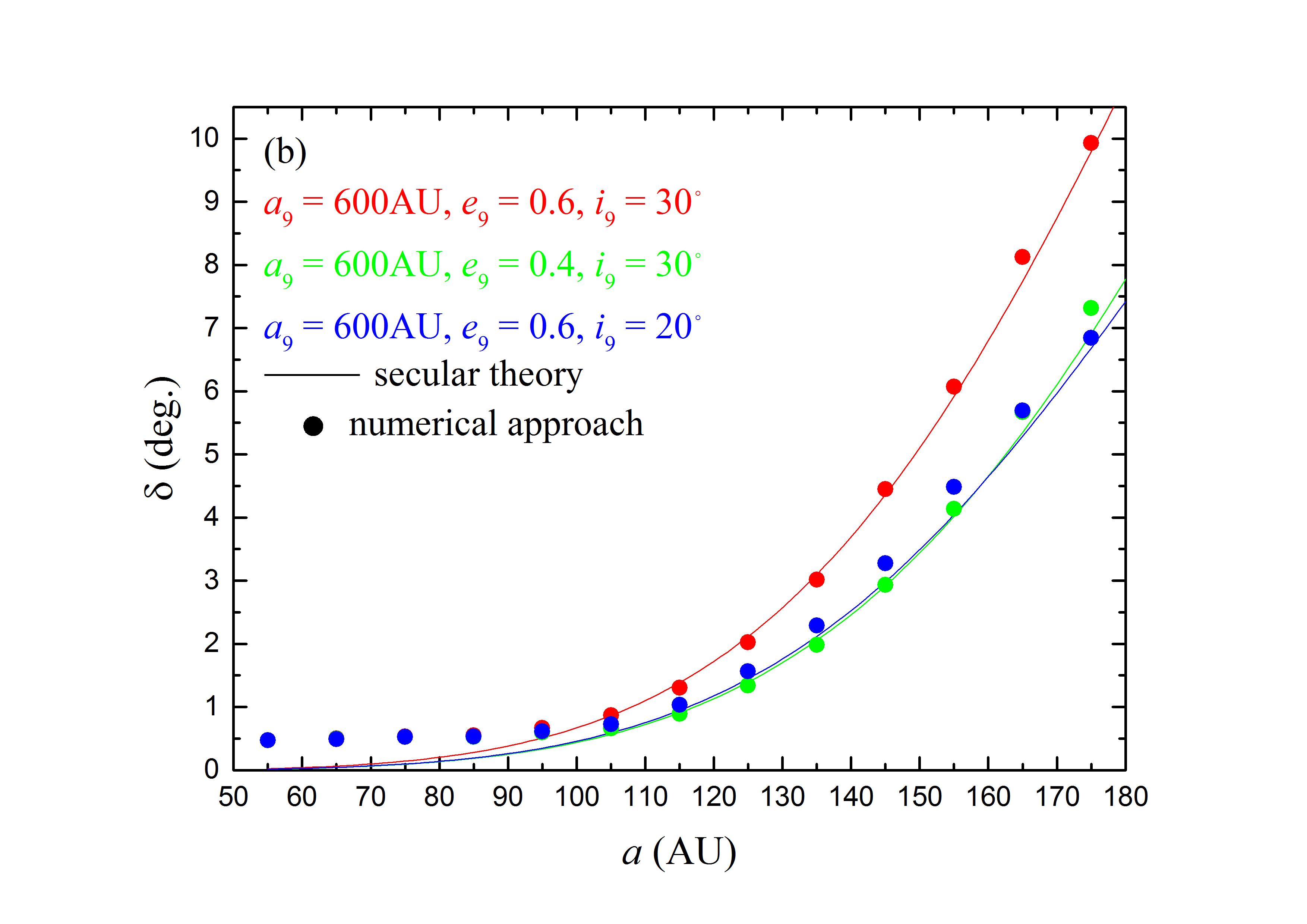}
  \end{minipage}
 \caption{Relative angle $\delta$ between the expected mean plane of the Kuiper belt and IP8 in the presence of a $10~m_{\oplus}$ Planet 9 with orbital elements $a_9$ (top panel: $=400$ AU; bottom panel: $=600$ AU), $e_9$ ($=0.4-0.6$), and $i_9$ ($=20^{\circ}-30^{\circ}$), as a function of the semimajor axis $a$ ($>50$ AU). The curve denotes the prediction from secular theory and the dots indicate the results from numerical computations with no approximations. The figure indicates these two approaches show good agreement.}
 \label{EMP}
\end{figure}

For the KBOs with instantaneous locations in three-dimensional space, the pole of the mean plane should be aligned with the forced inclination vector \citep{chia08}. Equation (\ref{approximate}) shows that the direction of $(q_0^1, p_0^1)$ depends on the secular mode of Planet 9 and varies with time. But as a magnitude, the forced inclination $i_0$ from Eq. (\ref{i0}) remains constant. Therefore, $i_0$ measures the relative angle $\delta$ between the mean plane of the Kuiper belt and IP8. Since the masses and semimajor axes of the known planets are constant in calculating $B$, at a certain $a$, the tilt of the Kuiper belt's mean plane (i.e., $\delta$) is completely determined by the mass ($m_9$) and orbital elements ($a_9$, $e_9$ and $i_9$) of Planet 9.

As described in the introduction, Planet 9 could possibly have $m_9=10~m_{\oplus}$, $a_9=400-600$ AU, $e_9=0.4-0.6$, and $i_9=20^{\circ}-30^{\circ}$. Within this parameter space, we calculated the tilt $\delta$ of the expected mean plane by our theoretical approach. Figure \ref{EMP} shows that the value of $\delta$ is as small as $<1^{\circ}$ in the semimajor axis range of $a\sim50-90$ AU, indicating that the Kuiper belt's mean plane should nearly coincide with IP8. But along with the increasing $a$, the mean plane would become more and more inclined relative to IP8, by up to the order of magnitude of $\delta\sim10^{\circ}$. As envisioned by \citet{volk17}, such a massive and distant Planet 9 has a negligible effect on the mean plane of the KBOs exterior to $a\sim100$ AU. We note that throughout this paper we do not consider the semimajor axis region of $q_9-a < 10$ AU ($q_9$ is the perihelion of Planet 9), where the KBOs could undergo strong gravitational interactions with Planet 9 and secular theory may be not available. 

\subsection{Numerical approach}
\label{numerical}

For the eight known planets, their masses, initial positions, and velocities are adopted from DE405 with epoch 1969 June 28 \citep{stan98}. Then we calculate the total angular momentum vector $\vec{H}_8$ of these planets, by which the inclination and longitude of the ascending node of IP8 can be determined \citep{soua12}. Using the rotational transformation, we proceed to change the reference plane from the mean equator of J2000.0 to IP8. Finally, we introduce Planet 9 with $m_9$, $a_9$, $e_9$ and $i_9$ to our solar system model.  

Considering a set of $N$ test particles, each having the position $\vec{r}_j$ and velocity $\vec{v}_j$, the total (unit) angular momentum vector is given by
\begin{equation}
\vec{H}=\sum_{j=1}^{N}\vec{r}_i\times\vec{v}_i.
\end{equation}
Then the relative angle $\delta$ between the vectors $\vec{H}$ and $\vec{H}_8$ is computed by \citep{li2019}
\begin{equation}
\delta=\arccos\left(\frac{\vec{H}_8\cdot\vec{H}}{|\vec{H}_8|\cdot|\vec{H}|}\right).
\label{angle}
\end{equation}
As long as the sample size $N$ is large enough, this angle could represent the deviation of the mean plane of the particles from IP8 \citep{camb18}.

\subsubsection{Pre-runs}

Given $m_9=10~m_{\oplus}$, firstly we chose the orbital elements of Planet 9 to be $a_9=400$ AU, $e_9=0.6$ and $i_9=30^{\circ}$. In this case, Planet 9 has the minimum perihelion $q_9=160$ AU and the highest inclination according to the proposed parameter space \citep{bail16, baty19}. Consequently, such an additional planet would exert the strongest influence on the mean plane of the Kuiper belt.

The test particles were uniformly distributed in the $a$ space between 50 AU and 150 AU, where 10 AU interior to the perihelion of Planet 9. Within each $a$-bin of 10 AU (e.g., $a=50-60$ AU), there were 101 test particles with even separation $\Delta a=0.1$ AU. As for the ``nominal'' population, initial $e$-values were taken to be 0.01, and initial $i$-values were randomly sampled in the range $0^{\circ}-20^{\circ}$. The other three orbital elements were all chosen randomly between $0^{\circ}$ and $360^{\circ}$. For the sake of saving computation time, in all numerical simulations performed below, the four terrestrial plants are added to the Sun; and the angular momentum vector $\vec{H}_8$ is determined by the four Jovian planets. This simplicity has little influence on the measurement of the angle $\delta$. We then integrated the system consisting of the Sun, Jovian planets, Planet 9, and test particles over 4.5 Gyr.

We find that because of the perturbations from Planet 9, the direction of the vector $\vec{H}_8$ (i.e., the pole of IP8) is not steady but evolves with time. The precession of the inclination of IP8 with respect to its initial plane observed in our simulations is identical to that shown in Fig. 2 of \citet{bail16}. Such that, in the numerical approach, our reference plane is indeed the instantaneous IP8. Accompanying the evolution of $\vec{H}_8$, the total angular momentum vector $\vec{H}$ of test particles also keeps changing direction, but the relative angle $\delta$ between these two vectors would no longer remain around zero. 

As we expected, at the end of the integration, the angle $\delta$ increases monotonously with increasing heliocentric distance, from $\sim1^{\circ}$ for the $a=50-60$ AU bin to $\sim6^{\circ}$ for the $a=120-130$ AU bin. For test particles farther than $a=130$ AU, they experienced stronger perturbations from Planet 9 and a small fraction survived, thus a meaningful measurement of the mean plane cannot be reached. The associated $\delta$-value is determined later from high $a$-resolution simulations with a larger sample size. Aside from the nominal population of test particles, we also carried out several additional simulations by varying either the initial $e$ or $i$, for example, $i=0.01^{\circ}$, $e=0-0.2,$ and $e=0.2-0.4$. These different inputs all reproduce nearly the same outcomes. The independence of the angle $\delta$ on particles $e$ and $i$ is easy to understand because these two orbital parameters are not visible in the calculation of the forced inclination $i_0$ (equivalent to $\delta$), as presented in Section \ref{theory}.

\subsubsection{High $a$-resolution runs}

\begin{figure}
 \hspace{0 cm}
  \includegraphics[width=8.5cm]{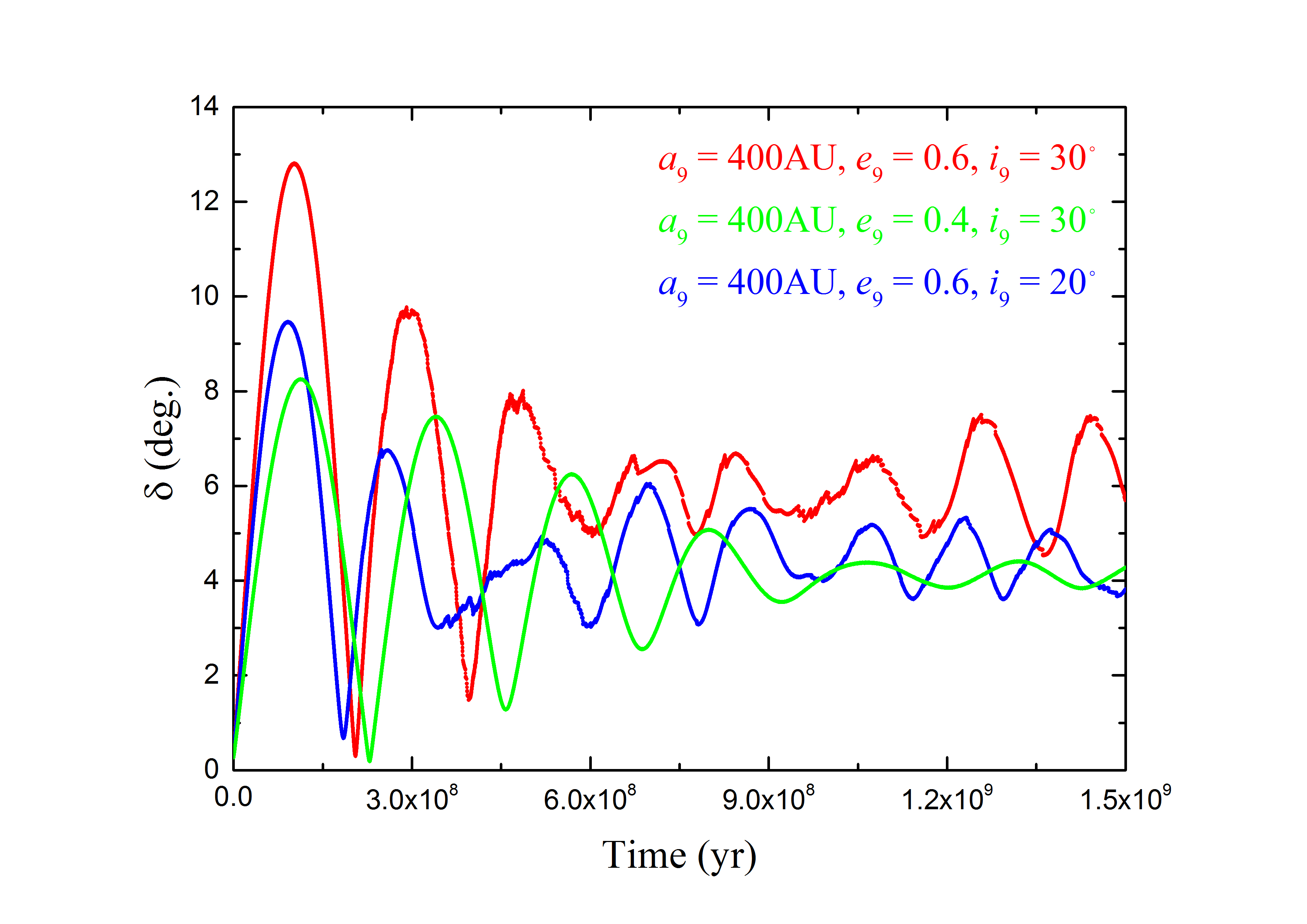}
  \caption{Time evolution of the angle $\delta$ for the semimajor axis bin of $a=120-130$ AU due to the presence of a $10~m_{\oplus}$ Planet 9 with different orbit sets $(a_9, e_9, i_9)$. The figure shows that $\delta$ can nearly converge to the average value at the end of the 1.5 Gyr integration.}
  \label{alphaT}
\end{figure}

In order to refine the value of $\delta$, we employed a higher spatial resolution of $\Delta a=0.02$ AU, which yields 501 test particles in each $a$-bin with a width of 10 AU (starting from $a=50-60$ AU). The initial eccentricities and inclinations were adopted to be the same as those of the nominal population mentioned about above, i.e., $e=0.01$ and $i=0^{\circ}-20^{\circ}$. Then we performed the high $a$-resolution runs with six different sets of $(a_9, e_9, i_9)$ for a $10~m_{\oplus}$ Planet 9, corresponding to the theoretical study in Section 2.1. In this subsection, we chose a shorter integration timescale of 1.5 Gyr for the numerical simulations. Even so, each run for a specific orbit set of Planet 9 would take over ten days of computing time on our workstation.

As an example, Fig. \ref{alphaT} shows the time evolution of the angle $\delta$ for the $a=120-130$ AU bin. We note that $\delta$ is osculating around the average value, and the amplitude could become rather small at the end of the integration. Thus for test particles in each semimajor axis bin of $[a^{(1)}, a^{(2)}]$, the angle $\delta$ is taken to be the average value during the integration, and it is assigned to the location of the median $a=(a^{(1)}+a^{(2)})/2$, as indicated by the dots in Fig. \ref{EMP}. In comparison with the angle $\delta$ predicted by the secular theory, as plotted by the curves in Fig. \ref{EMP}, we find that the numerical results are in good agreement. This can nicely support the validity of our theoretical approach developed in Section \ref{theory}. It should be noticed that, for the orbital elements of Planet 9, the theoretical approach only requires $a_9$, $e_9$, and $i_9$, but neither the argument of perihelion $\omega_9$ nor the longitude of ascending node $\Omega_9$. The parameters $\omega_9$ and $\Omega_9$ could control the orbital alignments of the 14 known KBOs with $a\ge250$ AU \citep{baty19}, but they should not affect the tilt of the Kuiper belt's mean plane.

Fig. \ref{EMP} shows that, within the $a<100$ AU region, both the theoretical and numerical values of $\delta$ are below $\sim1^{\circ}-2^{\circ}$. Considering the observational bias, we do not think such a slight warp in the Kuiper belt could be detected by future surveys. As a result, we focus on the KBOs between $100 < a < 200$ AU. Even so, this population is much closer and more detectable than the extremely distant KBOs with $a > 250$ AU that are believed to be clustered in physical space by now. We intend to use the angle $\delta$ as a constraint on the mass and orbit of the possible Planet 9 in the solar system.


\section{Kuiper belt observation}

\subsection{Real KBOs}
\label{realMP}

We selected the KBOs observed over multiple oppositions as of September 2019, taken from the Minor Planet Center\footnote{http://www.minorplanetcenter.net/iau/lists/TNOs.html}. Among these, there are 46 objects with semimajor axes $100 < a <200$ AU and perihelion distances $q>30$ AU. These objects comprise our sample KBOs and were used to calculate the mean plane of such a truncated Kuiper belt. In \citet{volk17}, the resonant KBOs were excluded since Neptune's mean motion resonances are not considered in secular theory. But in the direct N-body integrations performed in Section \ref{numerical}, we fully took the complete perturbations from the planets including Neptune into account. The agreement between our numerical and theoretical approaches suggests that the weak high-order resonances in the distant Kuiper belt could have little impact on the mean plane determination. As a matter of fact, \citet{sail17} found that, for the majority of resonant KBOs with $a>50$ AU, their secular behaviors can hardly be affected by mean motion resonances. Up to now, only three resonant KBOs have been identified in the $a>100$ AU region, i.e., 2004 PB$_{112}$ in the 5:27 resonance \citep{sail17}, 2015 KE$_{172}$, and 2007 TC$_{433}$ in the 1:9 resonance \citep{volk18}; these KBOs are in our sample. 

Since the number density of our sample KBOs is very low in the wide $a$-space, the mean plane measured by their total angular momentum may suffer from  severe observational bias. Instead, in this section we apply the alternative method employed in \citet{brow04} and \citet{volk17}: if a plane can, on average, go through all the sky-plane velocity vectors of the considered objects, then it defines the associated mean plane. For the unit pole vector $\vec{n}$ perpendicular to this mean plane, it can be computed by minimizing the residual \citep{volk17}
\begin{equation}
E=\sum_{i}|\vec{n}\cdot\vec{v_i}|,
\label{pole}
\end{equation}
where $\vec{v_i}$ is the unit vector of the sky-plane velocity of a KBO. All the $\vec{v_i}$  used were evaluated at a common epoch of 2019 April 27. Consequently, the mean plane can be achieved in a manner almost regardless of the discovery positions of the KBOs. Detailed descriptions can be found in the two papers cited above. It must be noted that when the number density of sample particles is large enough (e.g., that used in Section 2.2.2), we would obtain exactly the same mean plane by applying either the velocity vector (Eq. (\ref{pole})) or the angular momentum (Eq. (\ref{angle})), while the latter approach is much computationally cheaper.

We find that, for the $100 < a < 200$ AU Kuiper belt, using the directional velocity $\vec{v_i}$ to determine the mean plane yields an overall inclination $\tilde{\delta}\approx13.1^{\circ}$. This value seems too large to be authentic according to our theoretical results in Section 2. We then realized that by minimizing the residual $E$ in Eq. (\ref{pole}), such a large $\tilde{\delta}$ can result from the contamination of certain sample(s) with $\vec{v_i}$ deviated substantially from the others, especially when the sample size is very limited. By excluding a single KBO (2015 RQ$_{281}$), the angle $\tilde{\delta}$ can drop sharply to only $1.0^{\circ}$ and a comparable value could also be obtained even if we continue to remove some additional sample(s). With such a slight relative angle of $\tilde{\delta}\sim1.0^{\circ}$, the mean plane of the considered Kuiper belt could be regarded as nearly coincident with IP8. However, the census of the distant KBOs is far from observational completeness, thus the measurement error is clearly warranted.

\subsection{Monte Carlo samples}

For the uncertainty of the derived mean plane due to the small number of real observed samples, following \citet{li2019}, Monte Carlo simulations were constructed to estimate the possible $\tilde{\delta}$ for the overall $100 < a < 200$ AU Kuiper belt. It is obvious that the measurement error strongly depends on the space dispersion of the inclined KBOs. By adopting an unbiased inclination distribution from \citet{brow4b}, as
\begin{equation}
f(i)\propto \sin i \cdot \exp(-i^2/2\sigma^2),
\label{iDistribution}
\end{equation}
where the Gaussian standard deviation $\sigma=20^{\circ}$, we first created 100,000 synthetic samples with random inclinations $i$ relative to IP8. For each synthetic sample with assigned $i$, we randomly chose 
\begin{equation}
0.95a_{real} \le a \le1.05a_{real},
\label{aDistribution}
\end{equation}
and
\begin{equation}
0.95e_{real}\le e \le1.05e_{real},
\label{eDistribution}
\end{equation}
where $a_{real}$ is the semimajor axis of a random component from 568 real KBOs with multiple-opposition orbits and $a>50$ AU, and $e_{real}$ is the eccentricity of another random component \citep[also see][appendix C]{volk17}. The other three angles of an orbit were randomly selected in the range $0^{\circ}-360^{\circ}$. 

Then, for each object in the catalog of these real observed KBOs, near the latitude and longitude where each object was discovered, we searched for a corresponding synthetic sample. In this way we obtained a set of 568 synthetic objects, among which those with $100 < a <200$ AU and $q>30$ AU were selected to be our Monte Carlo samples. Subsequently, we measured the mean plane of a Monte Carlo population via Eq. (\ref{pole}). This procedure was repeated 10,000 times for the statistical analysis. 

We find that the 10,000 Monte Carlo populations give the angle $\tilde{\delta}=2.8^{\circ}\pm1.8^{\circ}$ with a $1\sigma$ confidence level. This result indicates that, for the overall $a=100-200$ AU range, the measured mean plane of the real KBOs (i.e., having $\tilde{\delta}\sim1.0^{\circ}$) could be deemed within $1\sigma$ error of the true mean plane. Accordingly, the mean plane of the Kuiper belt for this semimajor axis bin probably deviates from IP8 by less than $5^{\circ}$. This critical value could be served as an upper limit of $\tilde{\delta}$ to put constraints on the parameter space of Planet 9.

\section{Constraints for Planet 9}

With the addition of Planet 9 to the solar system, the expected mean plane of the Kuiper belt at semimajor axis $a>100$ AU would not remain in the vicinity of IP8, but can be substantially inclined by up to $\sim10^{\circ}$, as shown in Fig. \ref{EMP}. While based on the observational data of the KBOs, the true mean plane is found to have an inclination of $<5^{\circ}$ at $1\sigma$ confidence. We thereby suppose that the existence of Planet 9 is possible if the relative angle $\delta$ between the expected mean plane and IP8 is smaller than or comparable to $5^{\circ}$. The larger $\delta$ may indicate that Planet 9 has produced a distinguishable discrepancy from the current observation, leading to the unlikelihood of certain mass ($m_9$) or orbital elements ($a_9$, $e_9$, $i_9$). Since the theoretical and numerical approaches agree nicely  (see Fig. \ref{EMP}), we used the former to make our prediction, while the latter is too computationally expensive to fulfill an extensive suite of calculations with various parameters of Plane 9.

To examine the possibility of Planet 9, we consider a combination of the proposed $m_9$, $a_9$, $e_9$, and $i_9$, as already presented in the introduction. Then the forced inclination $i_0$ in Eq. (\ref{i0}), equivalent to the angle $\delta$, is solely a function of the location $a$ of the Kuiper belt. This is a specific calculation to every $a$, while the overall tilt of the expected mean plane for a wide semimajor axis range can be written as
\begin{equation}
\tilde{\delta}=\frac{\int_{a_{in}}^{a_{out}}\delta(a)\mbox{d}a}{{a_{in}}-{a_{out}}},
\label{cumulative_alpha}
\end{equation}
where ${a_{in}}=100$ AU and ${a_{out}}=200$ AU are the inner and outer edges of the considered Kuiper belt, respectively.

\begin{figure*}
  \centering
  \begin{minipage}[c]{1\textwidth}
  \vspace{0 cm}
  \includegraphics[width=9cm]{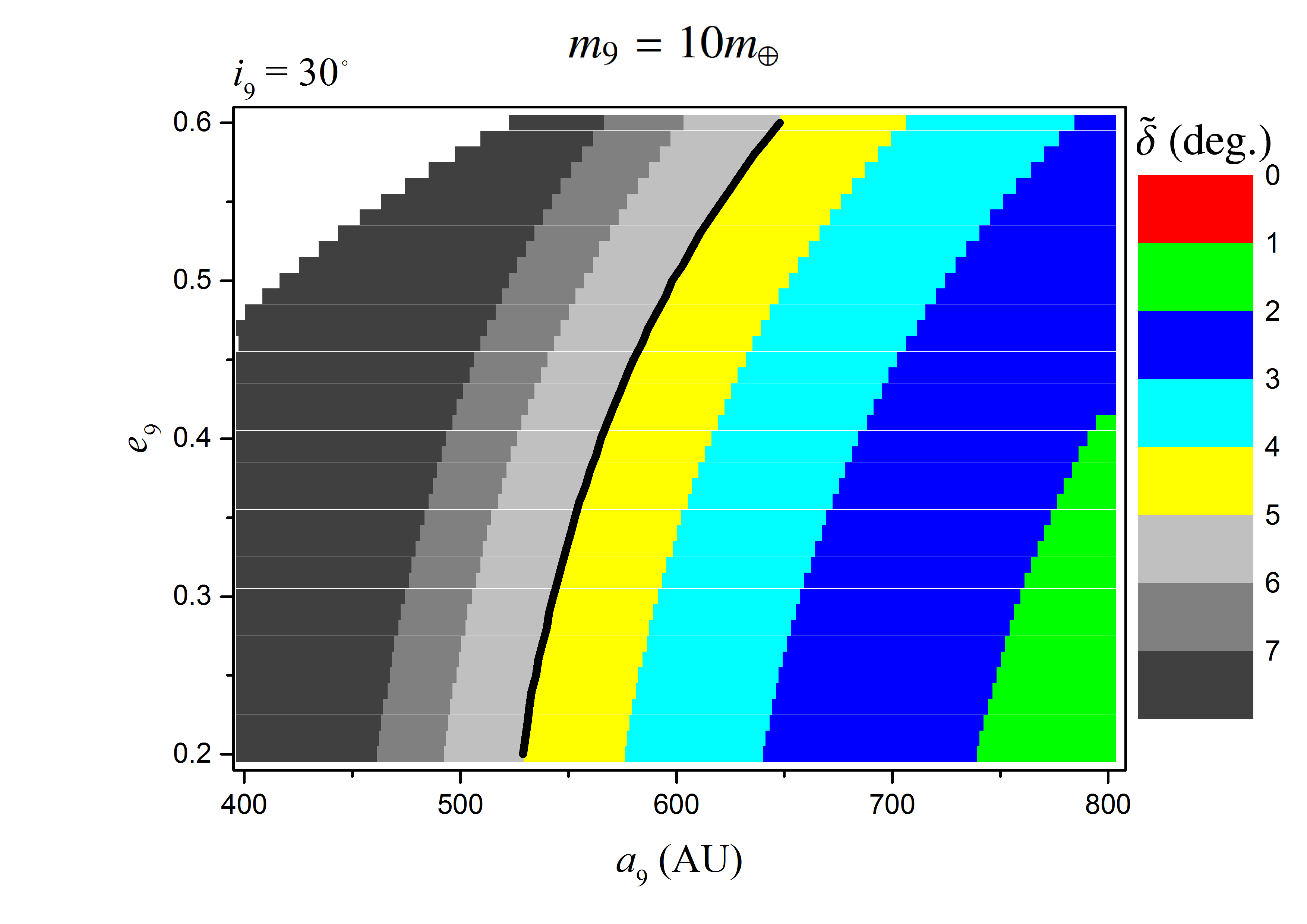}
  \includegraphics[width=9cm]{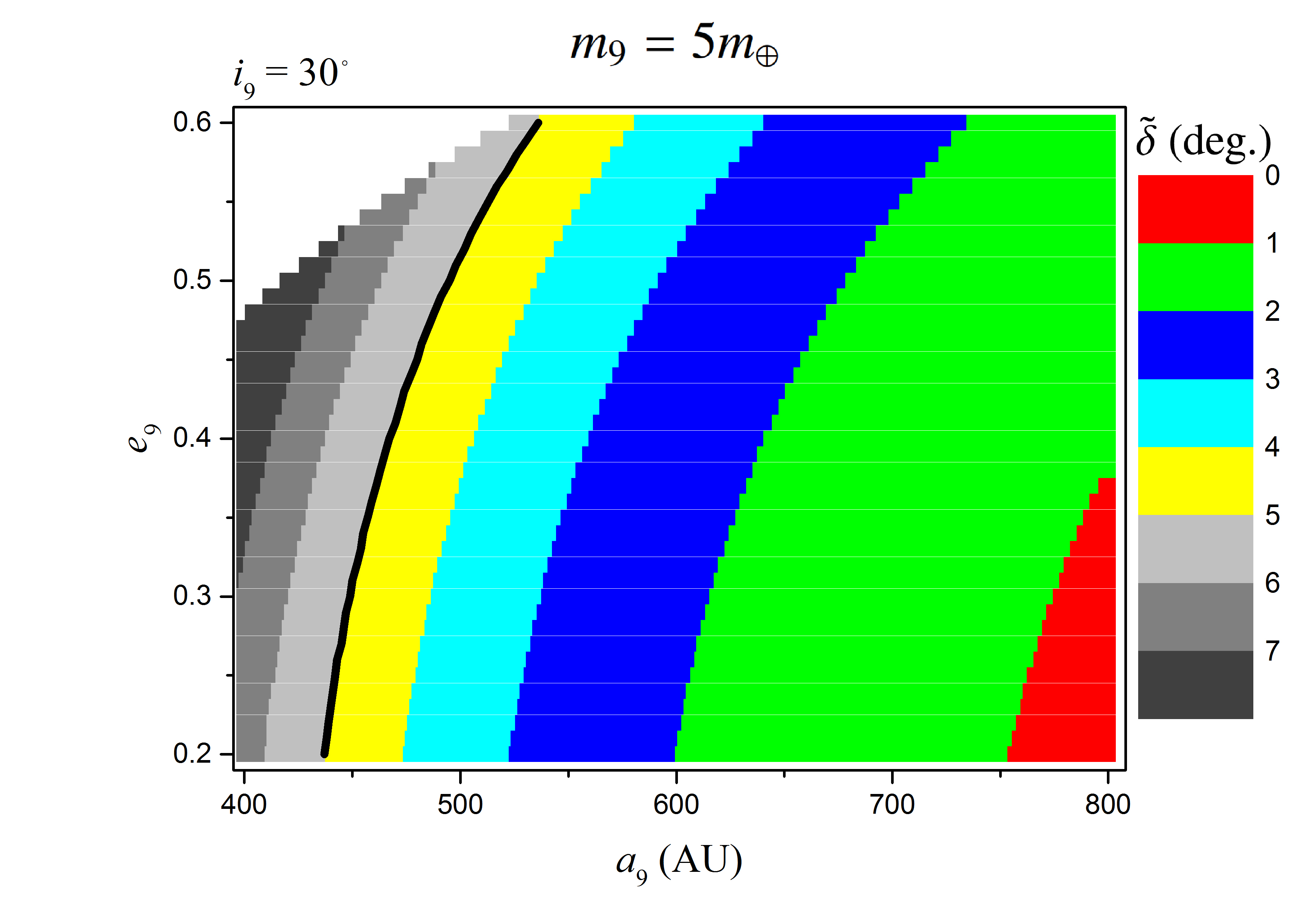}
  \end{minipage}
  \begin{minipage}[c]{1\textwidth}
  \vspace{-0.25 cm}
  \includegraphics[width=9cm]{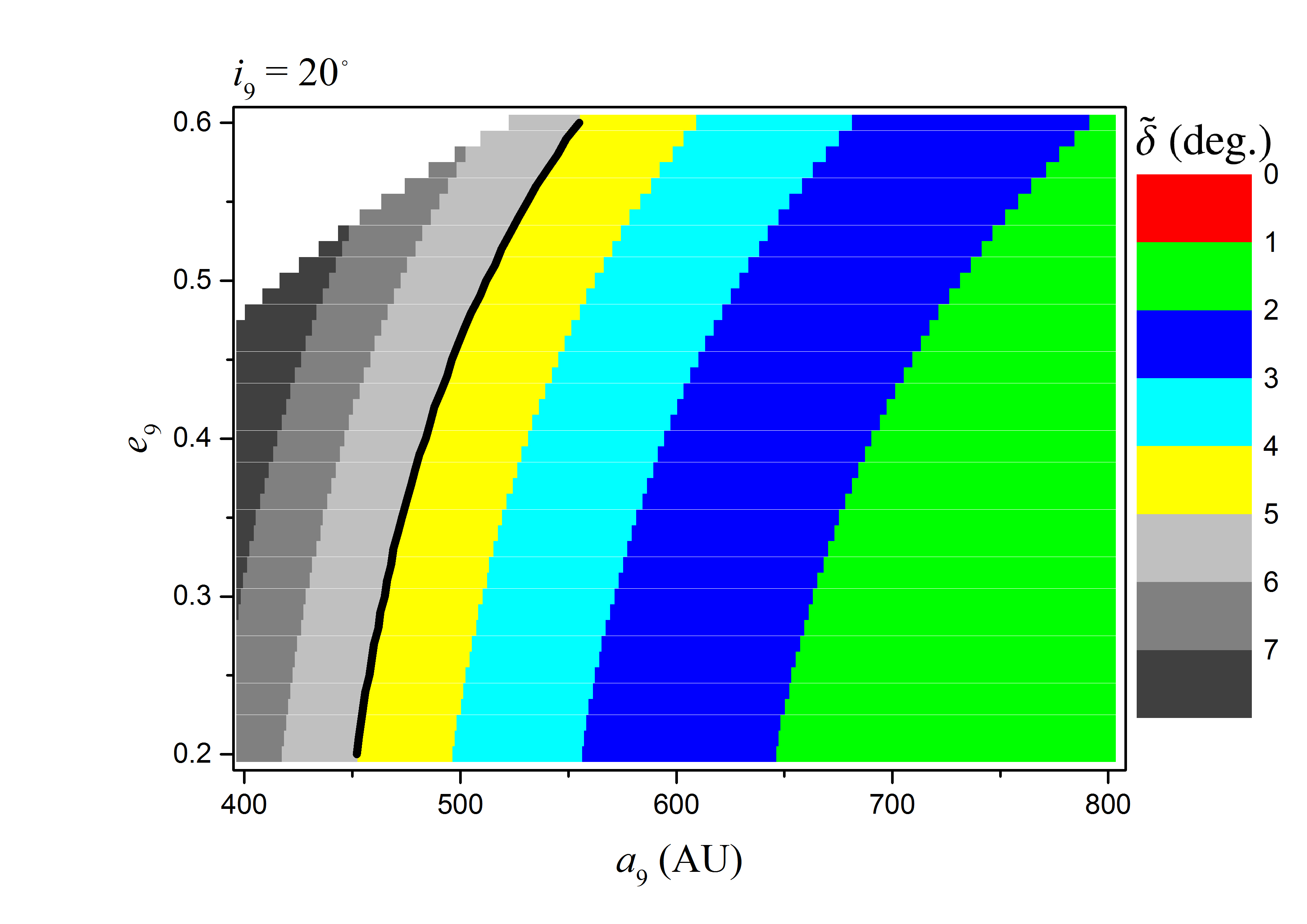}
  \includegraphics[width=9cm]{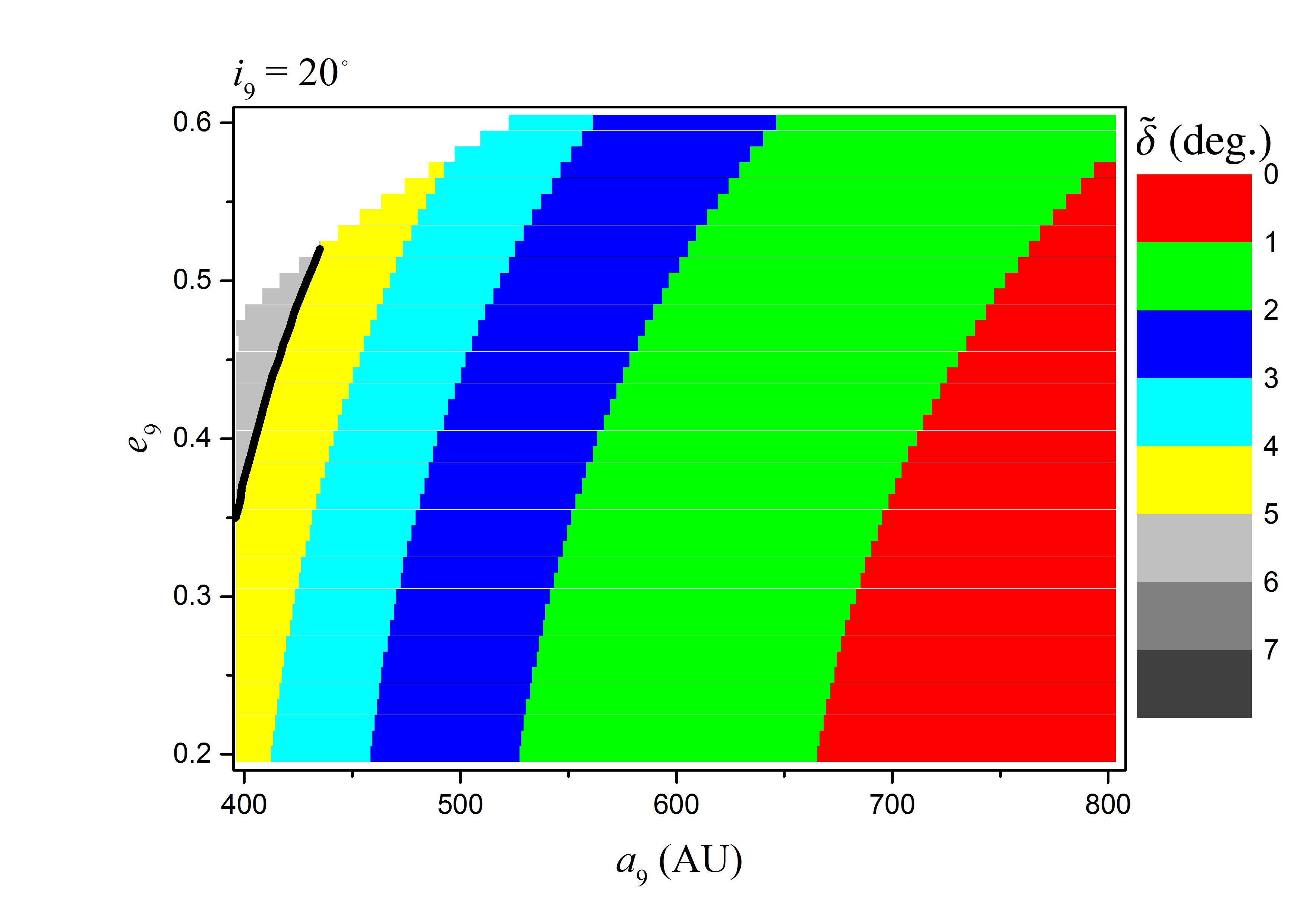}
  \end{minipage}
  \begin{minipage}[c]{1\textwidth}
  \vspace{-0.25 cm}
  \includegraphics[width=9cm]{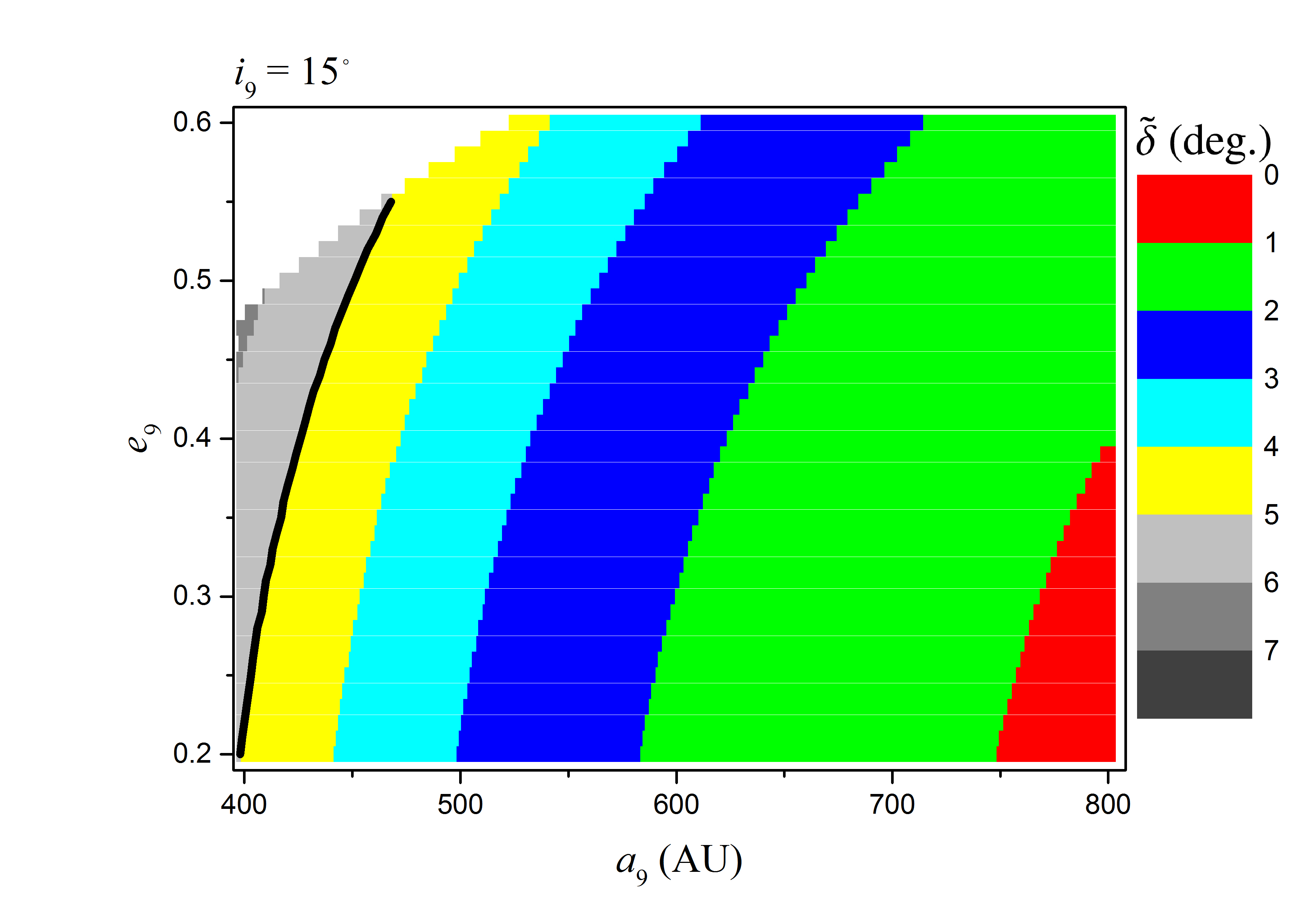}
  \includegraphics[width=9cm]{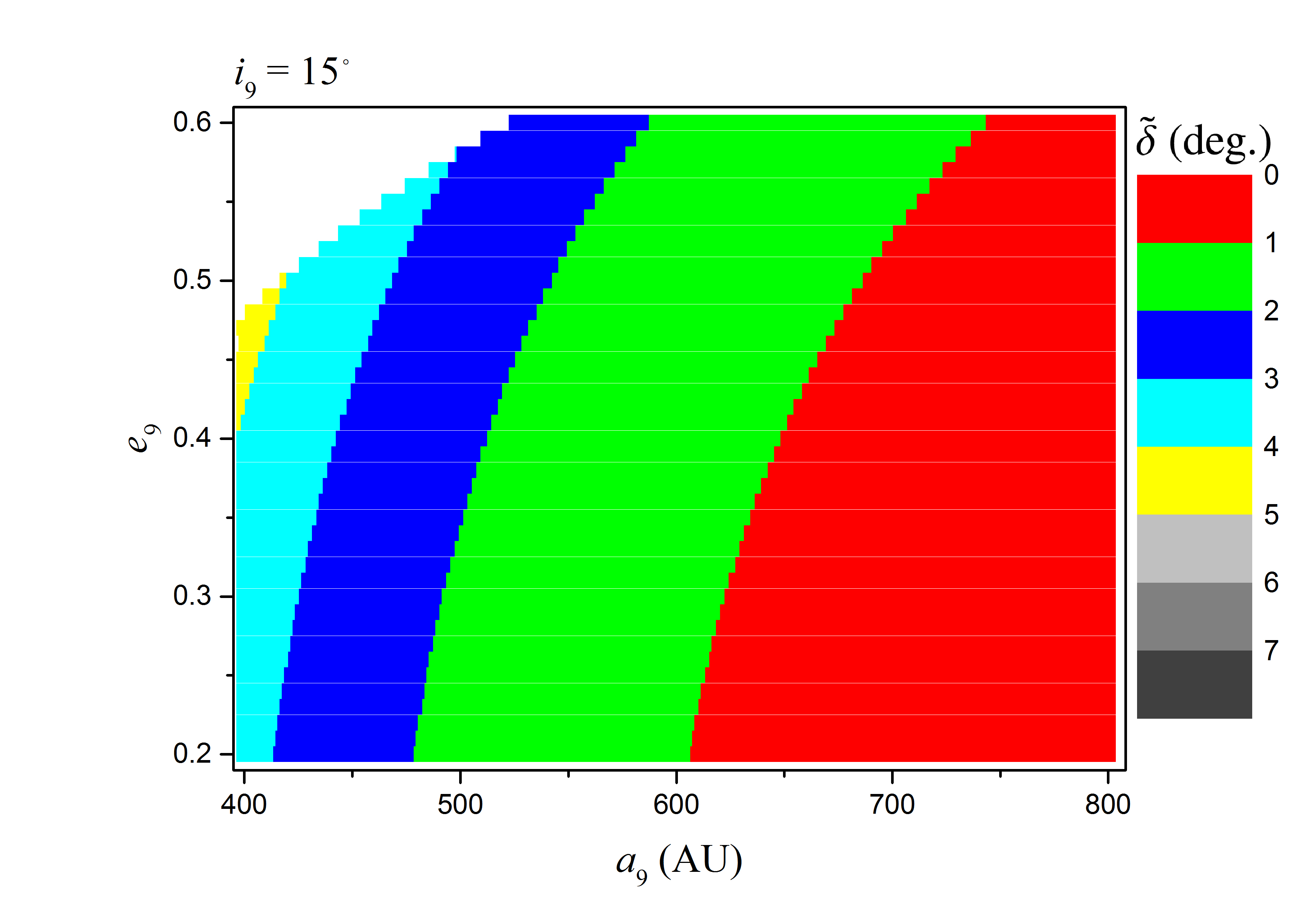}
  \end{minipage}
   \vspace{0 cm}
  \caption{Allowable ($a_9, e_9$) space for Planet 9 with $m_9=10~m_{\oplus}$ (left-hand column) and $5~m_{\oplus}$ (right-hand column), at representative inclinations $i_9=15^{\circ}$, $20^{\circ}$ and $30^{\circ}$. The color represents the relative angle $\tilde{\delta}$ between the expected mean plane of the $100 < a < 200$ AU Kuiper belt and IP8. The black curve demarcates the upper limit of $\tilde{\delta}=5^{\circ}$ deduced from the known KBOs: on the right side, the colorful regions with $\tilde{\delta}<5^{\circ}$ are the most probable zones for Planet 9; while on the left side, the gray regions with $\tilde{\delta}>5^{\circ}$ are less likely since the perturbation from Planet 9 seems too strong. The white regions indicate the zones with $q_9-a<10$ AU, where our theoretical approach for calculating $\tilde{\delta}$ may fail due to strong interactions between the KBOs and Planet 9.}
 \label{zones}
 \end{figure*}

In Fig. \ref{zones}, we plot the maps of the angle $\tilde{\delta}$ for several representative values of $i_9$. The left-hand and right-hand columns are for the cases of $m_9=10~m_{\oplus}$ and $5~m_{\oplus}$, respectively. On the right side of the black curve, the colorful regions refer to the relatively small deviation (i.e., $\tilde{\delta}<5^{\circ}$) of the mean plane of the Kuiper belt from IP8. This is allowable according to the results obtained in Section 3, thus Planet 9 with the given parameters is most probable. The gray regions on the left side of the black curve (i.e., $\tilde{\delta}>5^{\circ}$) are considered to be the less likely zones, and the darker the color the lower the possibility. It seems that the most inclined Planet 9 with $i_9=30^{\circ}$ possibly has an orbit of $a_9\gtrsim440-530$ AU (see top panels). We also notice that the unlikely gray regions have significantly shrunk with decreasing $i_9$  and the effective constraints on the $(a_9, e_9)$ pair can only be found for $i_9\gtrsim20^{\circ}$. In Fig. \ref{zones}, the white zone on the top left of each panel corresponds to the unconsidered region of $q_9-a<10$ AU, where the KBOs may experience chaotic evolution due to strong perturbations from Planet 9 and thus secular theory would be not applicable.

\section{Conclusions}

The existence of an inclined Planet 9 can induce IP8 to evolve from its initial plane, yielding the current tilt of $\sim6^{\circ}$ relative to the Sun's equator \citep{bail16, lai16, gome17}. As a natural extension, in this paper we explored the effect of this additional perturber on the mean plane of the Kuiper belt beyond 50 AU.

Firstly, in the context of the secular theory, we adopted and developed the theoretical approach of \citet{volk17} to determine the relative angle $\delta$ between the expected mean plane of the Kuiper belt and IP8  at every specific semimajor axis $a$. We found that in the region of $a=50-90$ AU, the expected mean plane nearly coincides with IP8. But at $a>100$ AU, a noticeable deviation between these two planes appears because $\delta$ could become as large as $\sim10^{\circ}$. By taking into account the complete perturbations from the known planets and Planet 9, we also constructed numerical simulations to compute the angle $\delta$ for test KBOs with considerable space dispersion. The good agreement validates our theoretical approach, which allows us to explore a large suite of the mass ($m_9$) and orbital elements ($a_9, e_9, m_9$) of Planet 9 within a reasonable amount of computing time.

Next, for the real KBOs with semimajor axes $100 < a <200$ AU, we obtained an overall mean plane deviating from IP8 by a small angle of $\tilde{\delta}\sim1.0^{\circ}$. Considering the small number of such distant KBO samples at present, we carried out Monte Carlo simulations to evaluate the measurement uncertainty due to the observational incompleteness. The results show that the measured $\tilde{\delta}$ is just within $1\sigma$ limit of $1.0^{\circ}-4.6^{\circ}$. We then suppose that an upper limit of $\tilde{\delta}$, taken to be $5^{\circ}$, can be used as a constraint on the parameter space of Planet 9.

By integrating the angle $\delta$ as a function of the semimajor axis $a$ in our theoretical approach, we are able to obtain the overall tilt $\tilde{\delta}$ of the expected mean plane for a wide range, i.e., $100 < a <200$ AU. In our final results for the proposed Planet 9 with $m_9=5-10~m_{\oplus}$ and $i_9=15^{\circ}-30^{\circ}$, we plot the maps of the angle $\tilde{\delta}$ on the ($a_9, e_9$) plane. Confined by the prescribed constraint of $\tilde{\delta}<5^{\circ}$, we propose that Planet 9 has the most probable orbit depicted by the colorful region in Fig. \ref{zones}:

(1) \underline{For $m_9=10~m_{\oplus}$:} Planet 9 could exist on a highly inclined orbit ($i_9=30^{\circ}$) in a more distant region beyond $a_9=530$ AU, or have moderate inclination $i_9=20^{\circ}$ ($15^{\circ}$) and smaller semimajor axis $a_9 > 450$ (400) AU.

(2) \underline{For $m_9=5~m_{\oplus}$:} Planet 9 is allowed to reside on a $i_9=30^{\circ}$ orbit with $a_9>440$ AU, or possibly any less inclined ($i_9<20^{\circ}$) orbit with $a_9>400$ AU. 

\noindent{With increasing $e_9$, the deduced minimum $a_9$ would grow slightly but not significantly; this critical $a_9$ value at $e_9=0.6$ is on a level of 1.2 times larger than the corresponding value at $e_9=0.2$.}

The above results could help to reduce the uncertainty of the proposed Planet 9's parameters. For instance, \citet{bail16} showed a $10~m_{\oplus}$ planet on a $a_9=400$ AU, $e_9=0.4-0.6$, $i_9=20^{\circ}-30^{\circ}$ orbit is capable of inducing the observed solar obliquity. However, such combinations of the mass and orbital elements clearly correspond to the unlikely regions shown in Fig. \ref{zones}. A less massive ($5~m_{\oplus}$) or more distant ($a_9=800$ AU) planet is suggested by \citet{baty19} from the clustering of the orbital planes of KBOs. Furthermore, \citet{kaib19} found that a much lower inclination ($i_9=5^{\circ}$) for Planet 9 seems favorable to replicate the inclination distribution of the observed scattering KBOs.

Since more and more faint KBOs will be discovered in the near future, for example, by the Large Synoptic Survey Telescope (LSST)\footnote{The LSST is a project of the US National Science Foundation, which has recently been renamed to the NSF Vera C. Rubin Observatory.} \citep{jone16}, they would help to further improve the measurement of the Kuiper belt's mean plane. Especially, for the discussed KBOs with $100<a<200$ AU, a larger number of samples will allow us to examine smaller $a$-bin so that a finer profile of the mean plane can be drawn. Consequently, even tighter constraints would be put on the mass and orbit of yet undiscovered Planet 9.

\begin{acknowledgements}
    
      This work was supported by the National Natural Science Foundation of China (Nos. 11973027 and 11933001), and National Key R\&D Program of China (2019YFA0706601). We would also like to express our sincere thanks to the anonymous referee for the valuable comments.
      
\end{acknowledgements}

%
%

\end{document}